%% file: paper.tex
\documentclass[conference]{IEEEtran}
\IEEEoverridecommandlockouts

\newif\ifcopyright

\copyrighttrue

\usepackage{cite}
\usepackage{amsmath,amssymb,amsfonts}
\usepackage{algorithmic}
\usepackage{graphicx}
\usepackage{textcomp}
\usepackage{xcolor}
\usepackage{xprintlen}
\usepackage{subcaption}
\usepackage{xfrac}

\input{include}

\usepackage[nolist,nohyperlinks]{acronym}
\begin{acronym}[TDMA]
  \acro{ARQ}{Automatic Repeat Request}
  \acro{ICS}{Industrial Control System}
  \acro{MitM}{Man-in-the-Middle}
  \acro{WSN}{Wireless Sensor Network}
  \acro{PER}{Packet Error Rate}
  \acro{CPS}{Cyber-physical System}
  \acro{DoS}{Denial of Service}
  \acro{IoT}{Internet of Things}
  \acro{MAC}{Message Authentication Code}
  \acro{HARQ}{Hybrid Automatic Repeat Request}
  \acro{FEC}{Forward Error Correction}
\end{acronym}

\newcommand{\agg}{\texttt{Agg$(\cdot)$}\xspace}
\newcommand{\comp}{\texttt{Comp$(\cdot)$}\xspace}
\newcommand{\sw}{\texttt{SW$(\cdot)$}\xspace}
\newcommand{\rd}{\texttt{R2D2$(\cdot)$}\xspace}

\definecolor{0bit}{HTML}{7A2531}
\definecolor{128bit}{HTML}{507E26}

\def\BibTeX{{\rm B\kern-.05em{\sc i\kern-.025em b}\kern-.08em
    T\kern-.1667em\lower.7ex\hbox{E}\kern-.125emX}}

\begin{document}

\title{MAC~Aggregation~over~Lossy~Channels~in~DTLS~1.3}


\author
{\rm Eric Wagner$^{\ast\dagger}$, David Heye$^{\dagger}$, Jan Bauer$^{\ast}$, Klaus Wehrle$^{\dagger}$, Martin Serror$^{\ast}$\\
    {\rm $^\ast$\textit{Cyber Analysis \& Defense}, Fraunhofer FKIE $\cdot$ \{firstname.lastname\}@fkie.fraunhofer.de } \\
    {\rm $^\dagger$\textit{Communication and Distributed Systems}, RWTH Aachen University $\cdot$ \{lastname\}@comsys.rwth-aachen.de}
}

\maketitle

\begin{abstract}
Aggregating Message Authentication Codes~(MACs) promises to save valuable bandwidth in resource-constrained environments.
The idea is simple: Instead of appending an authentication tag to each message in a communication stream, the integrity protection of multiple messages is aggregated into a single tag.
Recent studies postulate, \eg based on simulations, that these benefits also spread to wireless, and thus lossy, scenarios despite each lost packet typically resulting in the loss of integrity protection information for multiple messages.
In this paper, we investigate these claims in a real deployment.
Therefore, we first design a MAC aggregation extension for the Datagram Transport Layer Security~(DTLS)~1.3 protocol.
Afterward, we extensively evaluate the performance of MAC aggregation on a complete communication protocol stack on embedded hardware.
We find that MAC aggregation can indeed increase goodput by up to \SI{50}{\percent} and save up to \SI{17}{\percent} of energy expenditure for the transmission of short messages, even in lossy channels.
\end{abstract}

\acresetall

\begin{IEEEkeywords}
MAC aggregation, DTLS 1.3
\end{IEEEkeywords}

\section{Introduction}
\label{sec:introduction}

\ifcopyright
  \copyrightnotice{}
\fi

The (industrial) \ac{IoT} increasingly relies on wireless communication to interconnect sensors and actuators in a variety of domains, such as manufacturing, smart cities, or environment monitoring.
The limited capacity of the wireless transmission medium, shared among a growing number of devices, imposes stringent bandwidth constraints on these scenarios~\cite{2019_vitturi_industrial}.
Moreover, since \ac{IoT} devices are often battery-powered, minimizing communication as the primary driver of energy consumption is critical~\cite{2016_shaikh_radio_power}.

Consequently, optimizing the utilization of limited transmission resources remains an ongoing research challenge~\cite{2020_seferagic_survey}.
In this context, different streams of research explore techniques like compact protocol design~\cite{RFC9147,ctls_draft}, data compression~\cite{2012_raza_6lowpan}, data aggregation~\cite{2019_pushpalatha_energy}, or in-network processing~\cite{EEMP12} to reduce bandwidth consumption.
Furthermore, complementary approaches such as \ac{HARQ}~\cite{AAI+21} or relaying~\cite{HHCK07} aim at enhancing transmission reliability, thereby decreasing the need for retransmissions over lossy channels to further improve bandwidth efficiency.

At the same time, wireless channels amplify the need to secure transmitted messages as all communication is broadcast over a shared medium~\cite{2021_serror_challenges}.
To meet the generally recommended 128-bit security levels for integrity protection \cite{2020_nist}, each message must include a 16-byte authentication tag, known as a \ac{MAC}.
This overhead poses a significant challenge in bandwidth-constrained scenarios, where typical message sizes are only a few bytes~\cite{2024_wagner_when_and_how}. 
To mitigate this challenge, low-bandwidth protocols such as LoRaWAN~\cite{lora} and Sigfox~\cite{sigfox} reduce the security level by truncating authentication tags to 32~bits and 16–40~bits, respectively, well below the 64-bit minimum security level recommended by NIST already in 2016~\cite{nist-800-38B}.

Recent theoretical and simulative results highlight the promising potential of \ac{MAC} aggregation over bandwidth-constrained wireless channels~\cite{2024_wagner_when_and_how}.
In its simplest form, \ac{MAC} aggregation involves computing a joint authentication tag over $n$ consecutive messages, which is then appended only to the $n$-th message~\cite{2008_katz_aggregated-mac}.
Thus, the average overhead of integrity protection per message is reduced by sharing it among multiple messages.
On the downside, if any of these $n$ messages are lost, the integrity of none of the aggregated messages can be verified.
Despite this limitation, substantial goodput gains have been predicted for realistic wireless channels~\cite{2024_wagner_when_and_how}.

Nevertheless, before these promising results can be applied in the real world, two fundamental challenges must be solved.
First, we need to understand how MAC aggregation can be seamlessly integrated into existing communication protocols.
Here, we need to deal with delayed authentication as well as the selection of aggregation parameters, \eg the number of messages whose \acp{MAC} are aggregated.
Secondly, since simulations, as used by previous work, hardly reflect the dynamic channels seen in, \eg manufacturing environments, it is crucial to assess the true potential of \ac{MAC} aggregation in a real deployment.
Therefore, holistic evaluations of a complete communication protocol stack~(\eg to fully consider the overhead of each layer) under realistic channel behavior on actual hardware are required.

To the best of our knowledge, this paper is the first to investigate the applicability of MAC aggregation over lossy channels in a physical deployment.
Therefore, we integrate MAC aggregation into the Datagram Transport Layer Security~(DTLS)~1.3 protocol.
We decided to focus on DTLS~1.3 as it is a recent general-purpose security layer optimized for bandwidth-constrained networks, with the most compact header being only \SI{2}{byte} long.
Thus, DTLS~1.3 will likely be used in many future communication stacks for bandwidth-constrained communication in diverse domains and applications.
Nonetheless, our insights will also shed light on the expected potential of MAC aggregation into other protocols.

For our DTLS~1.3 integration, we define an extension for the DTLS~1.3 handshake to agree on using MAC aggregation.
This design enables the use of MAC aggregation with various existing and future cipher suites.
Moreover, we show how dynamic parameter updates allow adapting to changes in the transmission medium.
Finally, we offer an optional interface to retrieve received but not fully authenticated data optimistically to enable the vision of progressive authentication~\cite{2020_armknecht_promac}.
We show that \ac{MAC} aggregation seamlessly integrates into DTLS~1.3 without breaking backward compatibility and that goodput can thereby be significantly improved. 
Simultaneously, energy and bandwidth usage shrink in real-world deployments.

\textbf{Contributions.} To realize MAC aggregation for real-world deployments, we make the following contributions:
\begin{trianglelist}
  \item We design an extension to negotiate and use MAC aggregation in the DTLS 1.3 protocol.
  \item We develop a dynamic parameter selection scheme to adapt the MAC aggregation scheme to changing channels.
  \item We evaluate MAC aggregation in a real-world deployment and improve goodput by up to \SI{50}{\%} while cutting energy consumption by up to \SI{17}{\%}.
\end{trianglelist}

\section{Bandwidth Saving Concepts}
\label{sec:related-work}

Optimizing bandwidth utilization in resource-constrained wireless networks has been tackled from various angles.
In the following, we briefly overview different optimization concepts explored over the years. 
Afterward, we look at the different \ac{MAC} aggregation schemes that have been proposed recently.

\subsection{Alternative Approaches}

Efficient communication in resource-constrained wireless networks must carefully balance the trade-offs between bandwidth utilization, computational power, and energy consumption.
One approach to optimize bandwidth utilization is to reduce transmission overhead while maximizing the amount of information transmitted per packet~\cite{UPEl02}. 
This can be achieved, for example, through data compression and aggregation~\cite{KZM+21}, as well as by using piggybacking strategies, which include acknowledgments within data packets~\cite{RSMQ09}.

Other approaches aim to improve transmission reliability. 
\acf{HARQ}, for example, combines \ac{ARQ} with \ac{FEC} to enhance transmission reliability~\cite{AAI+21}.
The general idea is that the receiver stores corrupted packets and combines them with retransmitted versions to maximize the number of correctly decoded packets.
\ac{HARQ} can thus enhance reliability at the cost of increased end-to-end delay.

In a similar sense, cooperative communication~\cite{HHCK07} leverages spatial diversity by sharing transmission resources, \eg antennas, to enhance bandwidth efficiency.
This approach requires participating devices to coordinate and cooperate, \eg by relaying overheard messages to the intended receiver.
Relaying can significantly improve reliability depending on its implementation and use cases~\cite{2018_serror_relaying-eval}, which, in turn, reduces the number of retransmissions and lowers energy consumption.

However, incorporating security into resource-constrained wireless networks typically increases transmission overhead~\cite{LJPa20}.
A key challenge lies in mitigating the substantial overhead introduced by \acp{MAC}, particularly in narrow-bandwidth \ac{IoT} communication.
To address this, the following section reviews related work on \ac{MAC} aggregation techniques aimed at improving bandwidth efficiency.

\subsection{MAC Aggregation Schemes}

In recent years, various \ac{MAC} aggregation schemes have been proposed to split tags over multiple messages~\cite{2008_nilsson_compoundmac}, to progressively authenticate messages~\cite{2021_li_cumac, 2020_armknecht_promac, 2024_ginzboorg_authentication}, or to minimize the effect of packet loss~\cite{2022_wagner_spmac}.
These schemes are described by dependency sets, \ie which messages are protected with one tag, and an aggregation function, \ie how the tags are aggregated together.
We focus on two schemes, \agg and \rd, with XOR as their aggregation function and dependency sets that focus on different features.
\agg minimizes verification delays while \rd thwarts selective jamming attacks.
While we use XOR as the aggregation function due to its efficiency, other provable-secure aggregation functions, such as the one used by Whips~\cite{2020_armknecht_promac}, can also be considered.
While pseudorandom MACs, \eg HMAC, can be securely aggregated using XOR, some Carter-Wegman MAC constructions, \eg GMAC or Poly1305, are not provably secure when aggregated under XOR~\cite{2010_eikemeier_history}.
The schemes considered in the rest of this paper are detailed below:

\subsubsection*{\texttt{Trad}}
To quantify the performance of existing \ac{MAC} aggregation schemes, we compare them to the baseline performance of a generic traditional \ac{MAC} scheme.
Therefore, we consider a \ac{MAC} that authenticates each message $m_i$ with an individual tag $t_i$.
To achieve 128-bit security, this tag is \SI{16}{byte} long.

\subsubsection*{\texttt{Agg$(n)$}}
The most prominent aggregation scheme is aggregated MAC \agg as introduced in 2008~\cite{2008_katz_aggregated-mac} and also later extended to prevent reordering attacks~\cite{2010_eikemeier_history}, to allow messages to occur multiple times~\cite{2012_kolesnikov_multiplicity}, and to identify faulty messages in an aggregate~\cite{2018_hirose_non}.
For these schemes, an (aggregated) tag $t^{\text{agg}}$ is appended to every $n$-th message, where $n$ is the parameter for how many messages' authentication tags are aggregated.
For every $n$-th message, a tag is then computed by XOR-ing the authentication tags of all considered messages:
\begin{equation*}
  t_i^{\text{agg}} = \underset{i-n < k \leq i}{\bigoplus} t_{k} \quad \text{for } i\equiv -1\pmod{n}.
\end{equation*}

\subsubsection*{\texttt{R2D2$(n,o)$}}
The idea of progressive MACs is to protect messages with initially reduced security that is improved successively as more subsequent messages are received~\cite{2020_armknecht_promac, 2021_li_cumac,2024_ginzboorg_authentication, 2017_schmandt_minimac}.
Randomized and Resilient Dependency Distribution~(R2D2) improves upon this concept in the presence of packet loss by introducing dependency sets that bound the effect that a dropped packet can have on the verifiability of any other message~\cite{2022_wagner_spmac}.
If tags are \SI{2}{byte} long, any other message can lose at most 16~bits of security due to a lost message.
Furthermore, \rd randomizes the concrete dependency set $\mathcal{D}$ and assigns a different set to each bit of a tag to thwart selective jamming attacks.
The final aggregate tag $t^{\text{agg}}$ is thus a juxtaposition of bit-long tags and is defined as
\begin{equation*}
t^{\text{agg}}_i[j] = \bigoplus_{0\leq k<|\mathcal{D}_j|} t_{i-\mathcal{D}_j[k]}[k \cdot |t|+j]
\end{equation*}
with $\mathcal{D}_j[k]$ representing the $k$-th entry of $j$-th~bit's dependency set $\mathcal{D}_j$ and $t[k]$ represents the k-th bit in the tag $t$.
To provide full security under packet loss, \rd uses overprovisioning expressed as a factor $o$.
This factor defines, in percent, how much security may be extended beyond the target, \ie $o=100$ means that messages are protected by 256-bit security instead of the target of 128-bit security at the expense of longer tags.
We fix the number of aggregated tags $n$ to 8 as this value performs best according to prior experimentation~\cite{2024_wagner_when_and_how}.

\section{Integrating MAC Aggregation into DTLS~1.3}

Recent results suggest the benefits of MAC aggregation carry over to wireless scenarios, when bandwidth is constrained and messages are short~\cite{2024_wagner_when_and_how}.
To investigate this potential in a real-world setting, we set out to integrate MAC aggregation into DTLS~1.3~\cite{RFC9147}.
DTLS~1.3 specifically introduces many optimizations to save bandwidth in constrained environments and thus offers a suitable target for this investigation.
In the following, we present how we integrate MAC aggregation without breaking backward compatibility.

\subsection{Handshake Extension}

\begin{table}
  \begin{tabularx}{\columnwidth}{ p{1.3cm}<{\centering}p{2.6cm}<{\centering}p{3.5cm}<{\centering} }
  \textbf{Identifier} & \textbf{Dependency Set} & \textbf{Aggregation Function} \\
  \midrule
  \arrayrulecolor{lightgray}
  \code{0x00} & - & - \tabularnewline\midrule

      \code{0x01} & \agg & XOR \tabularnewline\midrule
      \arrayrulecolor{black}
      \code{0x02} & \rd & XOR \tabularnewline\midrule

  \end{tabularx}

  \caption{The two aggregation schemes as encoded in our proposed DTLS~1.3 extension. The identifiers \code{0x03} -- \code{0xff} are reserved for future schemes.}
  \label{tbl:schemes}
\end{table}

The DTLS~1.3 standard defines a handshake extension to implement new functionality.
We use this extension for clients to find out whether the server they connect to supports MAC aggregation and then to agree on a concrete aggregation scheme and parameters.
Therefore, we propose to add a new \code{ExtensionType} value of \eg \texttt{0x64}.
We define the structure of the extension messages as a sequence of aggregation schemes and respective parameter sets for both communication directions.
Each aggregation scheme is represented by a unique \SI{1}{\byte} identifier as shown in Table~\ref{tbl:schemes}.
Each aggregation scheme is defined by a dependency set and an aggregation function.

\begin{figure}
  \centering
  \begin{subfigure}[t]{\columnwidth}
    \centering
    \includegraphics[width=\columnwidth]{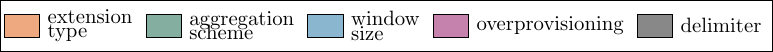}
  \end{subfigure}\\[.6em]
  \begin{subfigure}[t]{.48\columnwidth}
    \includegraphics[scale=.8]{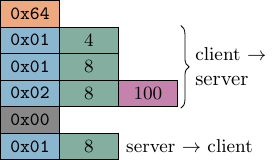}
    \caption{Extension request.}
    \label{fig:extension:request}
  \end{subfigure}
  \hfill
  \begin{subfigure}[t]{.48\columnwidth}
    \includegraphics[scale=.8]{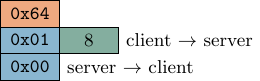}
    \caption{Extension response.}
    \label{fig:extension:response}
  \end{subfigure}
  \caption{During a DTLS~1.3 handshake, the client can append an extension to request MAC aggregation. The structure of this request is shown in Figure~\ref{fig:extension:request} and contains a list of supported aggregation schemes and parameters for both communication directions. The response, depicted in Figure~\ref{fig:extension:response}, selects the concrete schemes and parameters for the session. If no MAC aggregation is required, \code{0x00} is used, as in this response for server to client communication.}
  \label{fig:extension}
\end{figure}

The structure of the extensions is displayed in Figure~\ref{fig:extension}.
Each scheme is identified by its identifier and followed by either one or two aggregation parameters (\code{n}, \code{o}) depending on the aggregation scheme.
Supported aggregation schemes are defined for both directions individually, first from server to client and then from client to server, separated by a null byte. 

\begin{figure}[t]
  \centering
  \begin{subfigure}[b]{\columnwidth}
      \centering
      \includegraphics[width=\columnwidth, page=3]{figures/handshake/figure.pdf}
      \caption{If the server does not support the MAC aggregation extension, a normal DTLS~1.3 session is established.}
      \label{fig:handshake:without}
  \end{subfigure}
  \begin{subfigure}[b]{\columnwidth}
      \centering
      \includegraphics[width=\columnwidth, page=1]{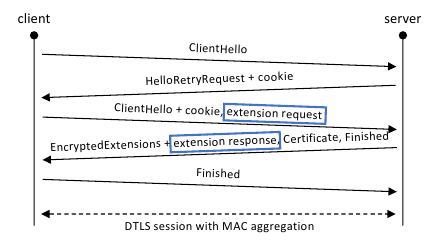}
      \caption{If the server does support the MAC aggregation extension, a MAC aggregation scheme is agreed upon and used in the established DTLS~1.3 to save bandwidth, conserve energy, and increase goodput.}
      \label{fig:handshake:with}
  \end{subfigure}
  \caption{
    The extension mechanism of DTLS~1.3 allows the seamless deployment of support for MAC aggregation alongside devices not supporting the extension.
  }
  \label{fig:handshake}
\end{figure}

The \textsc{Extension} is initially sent by the client to the server following the \textsc{ClientHello} to indicate which aggregation schemes it supports.
If the server does not know the \textsc{Extension}, the extension message is simply ignored as by the DTLS~1.3 standard~\cite{RFC9147} and a conventional DTLS~1.3 session is established as illustrated in Figure~\ref{fig:handshake:without}.

Otherwise, Figure~\ref{fig:handshake:with} shows how a DTLS~1.3 session with MAC aggregation is established.
Following the \textsc{ServerHello}, the server answers with an \textsc{EncryptedExtension} message that indicates support for MAC aggregation.
The answered \textsc{EncryptedExtension} contains exactly one aggregation scheme and parameter set for each transmission direction, selected from those advertised by the client.

The MAC algorithm and the desirable security levels for integrity protection are then defined by the negotiated cipher suite.
In the following, we assume that \texttt{TLS\_AES\_128\_GCM\_SHA256} is the agreed-upon cipher suite.
Thus, AES encryption in counter mode is used without expanding the ciphertext, 
and a GMAC is appended.

If the client and the server successfully negotiate MAC aggregation, all subsequent \code{record layer} messages, \ie \code{content type}~23, are protected with an aggregated tag.
However, control messages, such as \code{KeyUpdate} messages, still contain a full MAC such that they can be immediately and fully verified and processed.

\subsection{Record Layer Adaptations}


DTLS transfers data in \code{record layer} messages that contain a 16-byte authentication tag.
When using MAC aggregation, authentication tags are shorter than these \SI{16}{\byte}.
Consequently, authentication tags no longer have a constant length, and ciphertexts may be theoretically as short as \SI{2}{\byte}~(an encrypted \code{content type} with \SI{1}{\byte} payload and no authentication tag).
This optimization, however, leads to two challenges regarding the identification of content types of frames and the sequence number encryption method of DTLS~1.3.
In the following, we describe these two challenges as well as our proposed workarounds.

\subsubsection{Content Type Extraction}

Varying length authentication tags introduced by MAC aggregation make it challenging to identify non-record layer frames, such as \textsc{KeyUpdates}.
Such packets are sent with a full 16-byte tag to be immediately verifiable with full security and minimized reaction times.
However, to identify such packets, the \code{content type} must be decrypted and read.

In DTLS~1.3 ciphertexts, the \code{content type} is placed behind the payload and in front of the authentication tag.
As both of these fields may now vary in length, it is not easily possible to first extract the content type to know how to proceed with the message.
We could resort to always guessing a record layer protocol, and only if verification fails, consider that the message may be of another type with a longer authentication tag.
However, this solution gives an advantage to attackers as manipulations can no longer be clearly identified, and it increases overall processing.

\begin{figure}[t]
  \includegraphics[width=\columnwidth]{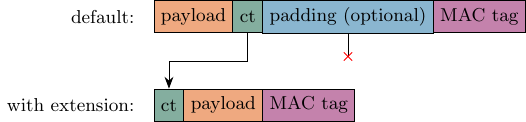}
  \caption{
      After agreeing on MAC aggregation, DTLS records carry the content type as the first field to make it easily recoverable during decryption. Consequently, MAC aggregation must be used in combination with a stream cipher, \eg AES in counter mode, as the content type no longer acts as delimited between the payload and padding.
      }
  \label{fig:ct}
\end{figure}

Instead, we reorder the layout of DTLS~1.3 ciphertexts after MAC aggregation has been agreed upon with the second \textsc{EncryptedExtension} message, as shown in Figure~\ref{fig:ct}.
Starting then, the content type of all encrypted messages is placed at the front of the ciphertext.
This is possible if we limit MAC aggregation to cipher suites that do not require padding, \eg those using AES in counter mode, as the \code{content type} currently also serves as padding delimiter.
Overall, this is only a minor limitation as the goal of MAC aggregation, \ie saving bandwidth, opposes the use of padding anyway.
After our layout changes, a receiver can decrypt the first byte (or block, depending on the selected mode of operation) to learn the alleged \code{content type}.
Based on this \code{content type}, the rest of the ciphertext can be processed.
The entire record can then be decrypted, and either the aggregated or the full authentication tag is verified.

\subsubsection{Sequence Number Encryption}

DTLS~1.3 encrypts the sequence number in its header to protect information from third-party observers.
However, because the plaintext sequence number is needed to ensure replay protection in the encryption of the payload, the sequence number is encrypted separately.
Here, a mask is computed by encrypting the first \qty{8}{\byte} or \SI{16}{\byte} of the ciphertext~(depending on whether a ChaCha or AES-based cipher suite is used) with a dedicated key.
This mask is then XOR-ed with the sequence number before transmission.
The sender recomputes this mask to reveal the plaintext sequence number before decrypting the message.

While the cipher suite typically ensures that the ciphertext is always longer than the  required \qty{8}{\byte} or \SI{16}{\byte}, MAC aggregation can lead to ciphertexts that are as short as \SI{2}{\byte}.
DTLS~1.3 proposes to pad too short plaintexts for sequence number encryption.
However, this would be counterproductive to the goal of MAC aggregation.
Instead, we propose to pad the input to the encryption algorithm.
First, we pad the input by the \SI{8}{\byte} of the expanded epoch number such that no evident collisions occur within different epochs.
If the input is still too short, the missing bytes are padded with null bytes.

Ultimately, this input to compute the mask for sequence number encryption has a lower entropy than if a longer ciphertext is available.
Hence, it may happen that the same mask is used twice during an epoch in a way that is obvious to outsiders.
Even in the worst case, with 1-byte payloads and \texttt{Agg(16)} as aggregation scheme, the likelihood of a first collision rises above \SI{50}{\percent} only after 274~messages. 
An observer listing in then still only learns the XOR-ed sequence number of both frames that used the same mask.
Only multiple such collisions allow an attacker to infer the sequence number.
However, an attacker who observes the channel for a prolonged period of time could also simply count the number of packets to gather the same information.
Hence, the real-world impact of reduced entropy for sequence number encryption is low, even in extreme scenarios with minimal payloads.

\begin{figure*}[t]
  \centering
  \begin{subfigure}[b]{.48\textwidth}
      \centering
      \includegraphics[width=\textwidth]{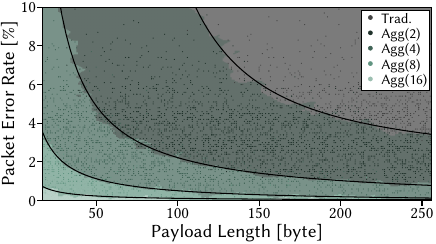}
      \caption{Aggressive MAC aggregation with \agg is only optimal for short payloads and with low \acsp{PER}.}
      \label{fig:param:agg}
  \end{subfigure}
  \hfill
  \begin{subfigure}[b]{.48\textwidth}
      \centering
      \includegraphics[width=\textwidth]{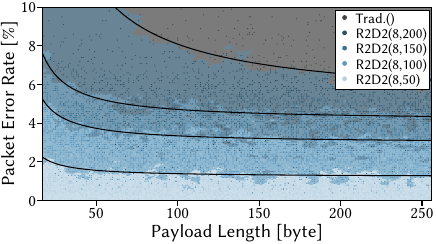}
      \caption{ \rd can aggregate more aggressively for longer payloads but it is rarely recommendable for \acsp{PER} above \SI{8}{\%}.}
      \label{fig:param:rd}
  \end{subfigure}
  \caption{We approximate the optimal aggregation parameters, \ie those yielding the highest goodput, based on boundaries described by formulas of the form $y = a\cdot e^{\frac{b}{x}}+c$, where $x$ is the payload lengths, $y$ is the current \ac{PER}, and $a$, $e$, $c$ as well as $b$ are parameters.}
  \label{fig:param}
\end{figure*}

\subsection{Upper Layer Interface}
\label{sec:design:interface}

In DTLS~1.3, received payload data is passed to the upper layer after its integrity has been verified.
Duplicate receptions are discarded.
Invalid messages (\eg due to an invalid authentication tag) are either silently discarded or responded to with a fatal alert.
For most situations, silently discarding invalid messages is recommended to minimize the need for expensive handshakes.
With MAC aggregation, most application data records cannot be fully verified upon reception.
Here, the receiving entity can decide how messages should be handed to the upper layer, either by buffering data until fully verified or by supplying data optimistically.

\subsubsection{Buffering Data Until Full Verification}

To match the behavior of conventional DTLS~1.3, unverified data could be buffered by the DTLS layer until its integrity is fully verified.
Application data is fully verified if a security level equivalent or higher to the baseline of the agreed-upon cipher suite is reached, \eg 128-bit security.

In this mode, all application data would be buffered until verified.
If a tag verification fails, the receiver can either silently discard the corresponding data or answer with a fatal error, depending on their configuration.
Moreover, buffered data is eventually discarded if data missing for verification is considered definitively lost.

\subsubsection{Optimistic Data Processing}

The idea of progressive message authentication is to optimistically process data before full integrity verification takes place~\cite{2020_armknecht_promac}.
The premise is that attacks are rare, and the benefits of low-latency data processing outweigh the cost of recovering from manipulated data that has evaded the lower security threshold.
Many scenarios, such as \acp{ICS}, are envisioned to benefit from such optimistic security~\cite{2009_szilagyi_flexible,2017_castellanos_retrofitting2}.
With MAC aggregation, such an optimistic operation can be supported.

Therefore, the receiver needs to set a security level threshold that determines when data is passed to the upper layer, and a callback function that is invoked if data integrity for already forwarded data is refuted retroactively.
This security threshold could \eg be set to 0, which would mean that all data is immediately passed to the upper layer unless tag verification explicitly fails.
Keep in mind that not all application data records carry a tag, or the carried tag cannot be verified due to lost preceding packets.
This threshold could also be set more conservatively to \eg 64-bit security, which corresponds to half the security provided by a full 16-byte tag and equals a chance of $1$ in $2^{64}$ that a manipulated message gets infiltrated. 

The callback function is called if authenticity is later refuted through the failed verification of a tag, and it provides the application layer with the number of processed bytes since the first malicious byte.
How to process such data and if this data invalidation should lead to a fatal error to close the communication channel is scenario-dependent.

\subsection{Dynamic Aggregation Parameters}
\label{sec:design:dynamic}

While the aggregation scheme is fixed for a session, one may want to dynamically adjust aggregation parameters if channel quality changes.
In extreme cases, a blocked line of sight could cause a high-reliability channel that usually benefits from aggressive aggregation to suddenly operate better without aggregation.
Hence, we devise a mechanism to update aggregation parameters dynamically.
This procedure is inspired by the DTLS~1.3 \textsc{KeyUpdate} mechanism but is always triggered by the receiver, who can judge the quality of the channel more accurately.

\subsubsection{Updating Aggregation Parameters}

To this end, we define a new (post-)handshake \textsc{AggregationUpdate} message, which either the client or the server may send during the session to request changing the window size and potentially other aggregation parameters at the peer.
Similar to other handshake messages without a built-in response, such as \textsc{KeyUpdate}, the \textsc{AggregationUpdate} is acknowledged by the peer upon reception, and it is retransmitted by the sender if the acknowledgment~(ACK) is not received before a timeout.
The \textsc{AggregationUpdate} is defined by a new message type and new parameters for the used aggregation scheme in the same format as in Figure~\ref{fig:extension}.

Upon sending the \textsc{AggregationUpdate}, the sender initializes a new decryption epoch with new decryption keys and the desired aggregation parameters.
Decryption keys are derived for a new epoch as if a \textsc{KeyUpdate} were received.
Since the \textsc{AggregationUpdate} is a handshake message, it contains a full authentication tag, and the receiver can verify the tag and process the message immediately upon reception.
All subsequent messages sent by the receiver must then use the new epoch for encryption. 
The decryption epoch at the receiver and the encryption epoch at the sender remain unchanged.

\subsubsection{Acknowledging \textsc{AggregationUpdate}s}

The last messages sent before receiving an \textsc{AggregationUpdate} may not be fully authenticated.
Each aggregation scheme should, therefore, define how to authenticate the last messages from the prior epoch after an \textsc{AggregationUpdate}.
One challenge is, however, that a receiver may not know how many messages were sent in the prior epoch.
Hence, to make the transition smoother, \textsc{Ack}s to \textsc{AggregationUpdate}s contain the sequence number of the last datagram from the previous epoch as an additional field.
If this \textsc{Ack} is lost, the receiver assumes that it received the last frame from the previous epoch.
If this were not the case, some tag verifications during the transition period may fail.
In that case, the unauthenticated data should be silently discarded. 

For \agg, the transition is taken care of by an additional tag carrying the aggregated authentication tag from all yet unauthenticated messages with the \textsc{Ack} to the \textsc{AggregationUpdate}.
If the scheme switches to traditional authentication with one full tag per message, the first frame of an epoch carries two tags, one for the current datagram and one for all not yet authenticated prior frames.

For \rd, all best-performing schemes depend on 8 prior messages and only differ by the overprovisioning factor.
To finalize an epoch to switch this parameter, we send a full authentication tag over the last 17 messages (half of the possible total verification delay) with the \textsc{Ack} to the \textsc{AggregationUpdate}.
Then, the next epoch, using full MACs or \rd with adapted overprovisioning, starts.

\subsubsection{Parameter Selection}
\label{sec:design:params}

Each receiver can decide when to switch parameters for the aggregation scheme based on which information it has, \eg to preempt a worsening or improving channel due to the receiver's movement.
In the following, we propose a generic reactive scheme to adapt to relatively long-term changes of the wireless link.

We want to understand how the schemes perform under different channel conditions.
Therefore, we generated binary packet loss traces for 10\,000 channels based on the Gilbert-Elliot model based on real-world measurements~\cite{2021_haenel_ge-params}.
For each of 10\,000 smapled traces with varying packet lengths, we check which parameters lead to the best goodput.
From these measurements, we observe that the best parameters mainly depend on the current payload length and packet error rate.
The respective best schemes are visualized as dots in Figure~\ref{fig:param}.

Then, we train a k-nearest neighbor~(KNN) classifier and use it to fill the area.
However, we want avoid porting a KNN classifier to often heavly resource-constrained devices that MAC aggregation aims to relieve.
Instead, we propose the following approximation of the KNN classifier.

The boundaries closely resemble the behavior of a function of the form $y = a\cdot e^{\frac{b}{x}}+c$, where $y$ is the \ac{PER} and $x$ is the payload length.
Hence, we performe a least-square fitting of the boundaries between parameters to obtain the parameters $a$, $b$, and $c$.
The resulting functions are plotted as black lines in Figure~\ref{fig:param}.
The receiver can then easily compute between which boundaries the current channel performs and adapt the parameterization accordingly.
To avoid unnecessary repeated switching, we propose the following mechanism to decide when to switch parameters.
Based on the running average \ac{PER} over the previous 200 packets, the receivers constantly assess if the currently selected parameters are optimal based on the curves identified in Figure~\ref{fig:param}.
If non-optimal parameters are selected for 50 consecutive \emph{transmitted} packets (the receiver counts lost packets once it identifies them through follow-up sequence numbers), the channel switches to the currently optimal parameters.
Thus, we efficiently determine and use optimal aggregation parameters for the current channel without burdening the channel with many parameter changes.

\begin{figure}[t]
  \includegraphics[width=\columnwidth]{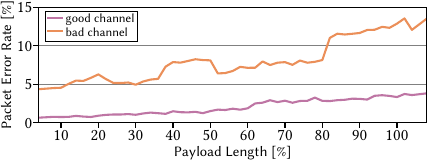}
  \caption{
      As expected, the \acp{PER} increase for longer payloads as the probability of uncorrectable bit flips increases.
  } 
  \label{fig:baseline}
\end{figure}

\begin{figure*}[t]
  \centering
  \begin{subfigure}{.48\textwidth}
      \centering
      \includegraphics[width=\textwidth]{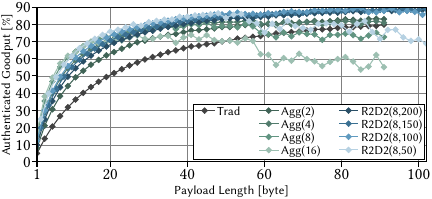}
      \caption{With low \acp{PER} on the good channel, most aggregation schemes achieve significant goodput improvements of, \eg \SI{32}{\%} for 30\,byte payloads.
      } 
      \label{fig:goodput:good}
  \end{subfigure}
  \hfill
  \begin{subfigure}{.48\textwidth}
      \centering
      \includegraphics[width=\textwidth]{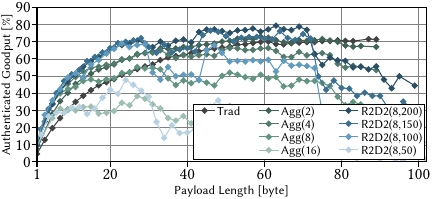}
      \caption{
        Even with higher \ac{PER} on the bad channel, an appropriate MAC aggregation scheme yields goodput improvements of up to
        \SI{146}{\%} for 1\,byte payloads and  \eg \SI{38}{\%} for 20\,byte payloads.}
      \label{fig:goodput:bad}
  \end{subfigure}
  \caption{MAC aggregation can significantly improve authenticated goodput but can decline if applied too aggressively, especially for channels with higher \acp{PER}. On the bad channel, for example, even conservative MAC aggregation falls short of appending one long MAC to each message for payloads longer than \SI{80}{byte}.
  }
  \label{fig:goodput}
\end{figure*}

\begin{figure*}[t]
  \centering
  \begin{subfigure}{.48\textwidth}
      \centering
      \includegraphics[width=\textwidth]{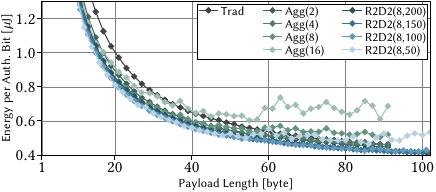}
      \caption{MAC aggregation on the good channel with a low \ac{PER} even yields \SI{8.4}{\%} energy saving for the longest supported payload.}
      \label{fig:energy:good}
  \end{subfigure}
  \hfill
  \begin{subfigure}{.48\textwidth}
      \centering
      \includegraphics[width=\textwidth]{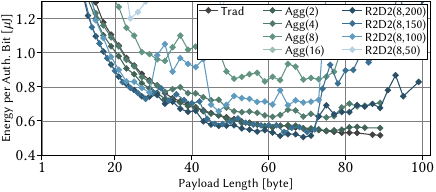}
      \caption{
        For short payloads, more conservative MAC aggregation can save up to \SI{20.6}{\%} of energy even with high \ac{PER} on the bad channel.
      }
      \label{fig:energy:bad}
  \end{subfigure}
  \caption{In many scenarios, MAC aggregation can lead to more than \SI{10}{\%} energy savings. However, the scheme and parameters must be properly chosen, especially on channels with higher \acp{PER}.}
  \label{fig:energy}
\end{figure*}

\section{Performance Evaluation}

To assess MAC aggregation in realistic DTLS~1.3 scenarios, we perform a variety of tests. 
First, we evaluate bandwidth and energy saving for different payload lengths.
Then, we dive deeper into the performance of MAC aggregation based on concrete scenarios, assessing goodput improvements, authentication delays, and dynamic aggregation parameter selection.

\subsection{Measurement Setup}

We conduct our measurements on two Zolertia RE-Mote boards equipped with a Cortex\,M3\,@\,32\,MHz, 32-bit CPU.
We use the Contiki-NG operating system and adapt the wolfSSL DTLS~1.3 implementation to support MAC aggregation.
The measurements are conducted in an over \SI{1000}{\square\meter} test and experimentation lab for energy-related equipment and components.
For most measurements, the client and servers are mounted on a fence at a height of \SI{1.94}{m} and \SI{12.50}{m} apart.
Different channel quality is achieved by changing the orientation of the antenna.
Interference was caused mainly by several 802.11~networks deployed in the facility.
When a stable channel was desired for comparability, measurements were conducted during the night or over the weekend, when network activity was severely reduced.

\subsection{Aggregated MACs vs. Longer Packets}
\label{sec:eval:estimation}

First, we assess MAC aggregation in DTLS~1.3 for a broad range of scenarios.
Therefore, we collect network traces for IEEE 802.15.4 packets with payload lengths between \SI{4}{\byte} and \SI{108}{\byte} over a good and a bad channel.
We plot the resulting \ac{PER} in Figure~\ref{fig:baseline}. 
As expected, the \ac{PER} increases for larger payloads.
It varies between \SI{0.65}{\%} and \SI{3.81}{\%} on the good channel, and between \SI{4.35}{\%} and \SI{13.52}{\%} on the bad channel.
We stitch these traces together according to the DTLS~1.3 packet sizes required by different MAC aggregation schemes for fixed payload sizes.
This method approximates the performance of MAC aggregation for a wider range of scenarios than possible if measuring the scenarios individually.

\subsubsection{Goodput Improvements}

We first investigate the potential \emph{authenticated goodput} improvements achieved by MAC aggregation.
We define authenticated goodput as the ratio between received fully authenticated payload bytes and the overall transmitted bytes.
We compare the authenticated goodput by the different schemes and traditional MACs~(without aggregation) for different payload lengths in Figure~\ref{fig:goodput}.

On the good channel in Figure~\ref{fig:goodput:good}, we see that MAC aggregation significantly improves authenticated goodput.
The most significant gains are observed for short payloads, which is expected as the overhead of MACs is proportionally larger.
We also see that the two schemes \agg and \rd perform similarly with different parameters.
While the most aggressive aggregations, \code{Agg(16)} and \code{R2D2(8,50)}, perform best for short messages, they also decline quickly with larger payloads due to increased  packet loss~(\cf\,Fig.~\ref{fig:baseline}).
Slightly more conservative aggregation outperforms traditional MACs over the full range of 802.15.4 payload sizes. 
Moreover, \rd reduces the maximum occupied space by authentication data in a DTLS~1.3 frame, thus enabling support for longer payloads sent within individual packets (\ie \SI{91}{\byte} vs. up to \SI{104}{\byte}).

On the bad channel in Figure~\ref{fig:goodput:bad}, we observe that aggressive aggregations, \eg \code{Agg(16)} and \code{R2D2(8,50)}, is not effective.
Still, more conservative aggregation, especially with \rd's resilience to packet loss, outperforms traditional MACs up until 72-byte payloads.
This falloff coincides with the channels \ac{PER} raising over \SI{8}{\%}, above which \rd rarely outperforms traditional MACs as noted in Section~\ref{sec:design:params}.
Overall, we still see, for example, a \SI{38}{\%} goodput improvement by \code{R2D2(8,200)} for 20-byte payloads.

We can thus conclude that MAC aggregation does indeed improve authenticated  goodput.
However, if the aggregation is too aggressive for a given channel, goodput quickly decreases.

\begin{figure*}[t]
  \includegraphics[width=\textwidth]{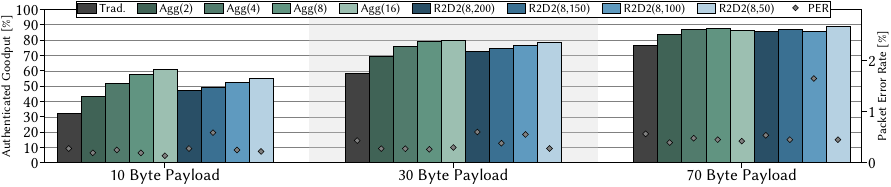}
  \caption{
    Evaluating concrete scenarios, we again see that short payloads benefit most from MAC aggregation with authenticated goodput improvements of up to \SI{90.6}{\%}, \SI{36.5}{\%}, and \SI{16.6}{\%} for 10-byte, 30-byte, and 70-byte payloads, respectively.
  } 
  \label{fig:static:goodput}
\end{figure*}

\begin{table}
  \centering
  \renewcommand{\arraystretch}{1.1}
  \begin{tabularx}{7.5cm}{ p{2cm}<{\centering}|p{2cm}<{\centering}|p{2cm}<{\centering} }
    \textbf{Payload} & \textbf{Frequency} & \textbf{Median PER} \\
    \midrule
    \arrayrulecolor{lightgray}
    \SI{10}{B}  & \SI{1}{Hz} & \SI{0.25}{\%} \tabularnewline\midrule
    \SI{30}{B} & \SI{10}{Hz} & \SI{0.30}{\%} \tabularnewline\midrule
    \arrayrulecolor{black}
    \SI{70}{B} & \SI{1}{Hz} & \SI{0.45}{\%} \tabularnewline
    \midrule
  \end{tabularx}
  \caption{Configurations of the three static scenarios.}
  \vspace*{-5mm}
  \label{tab:scenarios}
\end{table}

\subsubsection{Energy Consumption}

In resource-constrained scenarios, the longevity of battery-operated devices may also be a critical concern.
To derive the energy cost in Figure~\ref{fig:energy}, we first measure the energy cost of transmitting packets of varying sizes, where we observe an expected linear increase from around \SI{0.6}{\micro\joule} for 4-byte packets to around \SI{2.6}{\micro\joule} for 110-byte packets.

Figure~\ref{fig:energy:good} shows that MAC aggregation saves energy for all payload lengths on the good channel.
The best improvements occur for small payloads, with \eg energy saving of \SI{27.1}{\%} for 10-byte payloads by \code{R2D2(8,50)}.
But also the energy cost for longer payloads is reduced.
For example, \code{R2D2(8,150)} reduces costs by \SI{8.4}{\%} from \qty{0.46}{\micro\joule} to \qty{0.42}{\micro\joule} for the maximum supported 91-byte payload.
These savings can be extended to \SI{10.1}{\%} per authenticated bit when considering that \code{R2D2(8,150)} supports up to 102-byte payloads.

On the bad channel in Figure~\ref{fig:energy:bad}, we see that energy saving closely correlates with goodput improvement.
Here, we see energy savings of up to \SI{20.6}{\%} for 21-byte payloads with \code{R2D2(8,150)}.
However, for longer payloads above \SI{70}{\byte}, even \code{Agg(2)} is outperformed by traditional MACs.

MAC aggregation leads to energy saving unless payloads are long and \acp{PER} are high.
Therefore, the aggregation scheme and its parameters must, however, be appropriately chosen.

\subsection{Static Aggregation Parameters}
\label{sec:eval:static}

Now, we examine the performance of MAC aggregation based on three concrete scenarios.
We analyze the achieved goodput and compare it to the results of Figure~\ref{fig:goodput:good} before assessing the authentication delays of the different schemes.

\subsubsection{Setup}
To cover a variety of potential scenarios, we send 10-byte and 70-byte payloads with a frequency of \SI{1}{Hz} (\ie one packet per second) and 30-byte packets with a frequency of \SI{10}{Hz}, as summarized in Table~\ref{tab:scenarios}.
For each of the communication scenarios, we compare the performance of traditional 16-byte MACs to the same MAC aggregation schemes as before.
For each measurement of the nine schemes, our evaluation ran for \SI{1}{hour}, \ie a total of 27~hours, over a quiet weekend.
The grey diamonds on the second y-axis in Figure~\ref{fig:static:goodput} show the minor variations in PERs that can be explained by varying packet sizes, \eg scenarios with traditional 16-byte MACs yield slightly higher \acp{PER} due to the longer header.
The outliers, \eg \code{R2D2(8,50)} for 70-byte payloads, are caused by a short burst of high packet loss.

\begin{figure}[t]
  \centering
  \begin{subfigure}[t]{\columnwidth}
    \centering
    \includegraphics{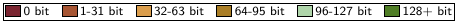}
  \end{subfigure}
  \begin{subfigure}[t]{.49\columnwidth}
    \includegraphics[width=\textwidth]{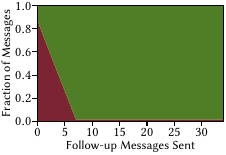}
    \caption{\code{Agg(8)}}
    \label{fig:poc:a}
  \end{subfigure}
  \hfill
  \begin{subfigure}[t]{.49\columnwidth}
    \includegraphics[width=\textwidth]{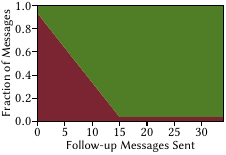}
    \caption{\code{Agg(16)}}
    \label{fig:poc:b}
  \end{subfigure}
  \hfill
  \begin{subfigure}[t]{.49\columnwidth}
    \includegraphics[width=\textwidth]{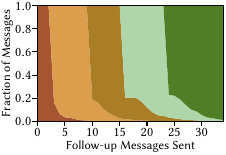}
    \caption{\code{R2D2(8,100)}}
    \label{fig:poc:c}
  \end{subfigure}
  \hfill
  \begin{subfigure}[t]{.49\columnwidth}
    \includegraphics[width=\textwidth]{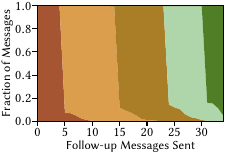}
    \caption{\code{R2D2(8,50)}}
    \label{fig:poc:d}
  \end{subfigure}

  \caption{ \agg provides quicker full authenticity but some data never receives any security guarantees. Meanwhile, \rd provides progressive authenticity improvements.}
  \vspace{-5mm}
  \label{fig:poc}
\end{figure}

\begin{figure*}
  \centering

  \begin{subfigure}[t]{\textwidth}
    \includegraphics[width=\textwidth]{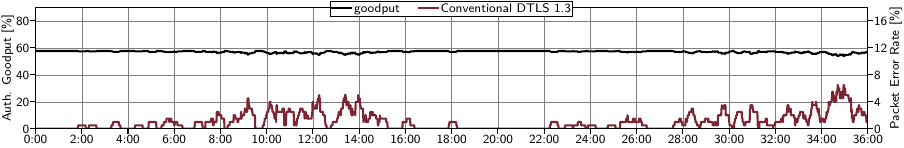}
    \vspace*{-6mm}
    \caption{The conventional DTLS~1.3 connection achieves an overall authenticated goodput of \SI{57.0}{\%}.}
    \label{fig:dynamic:vanilla}
  \end{subfigure}
  \par\medskip
  \begin{subfigure}[t]{\textwidth}
    \includegraphics[width=\textwidth]{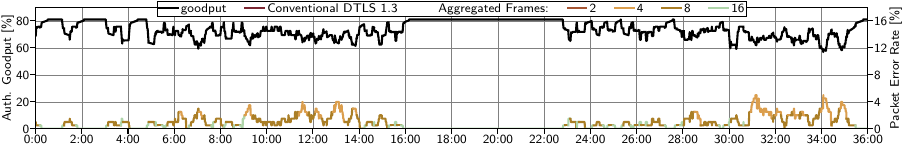}
    \vspace*{-6mm}
    \caption{The DTLS~1.3 with dynamic MAC aggregtion based on \agg achieves an overall authenticated goodput of \SI{74.4}{\%}.}
    \label{fig:dynamic:agg}
  \end{subfigure}
  \par\medskip
  \begin{subfigure}[t]{\textwidth}
    \includegraphics[width=\textwidth]{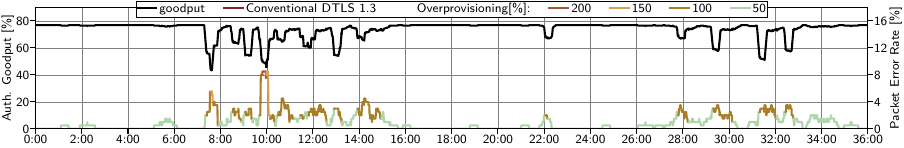}
    \vspace*{-6mm}
    \caption{The DTLS~1.3 with dynamic MAC aggregtion based on \rd achieves an overall authenticated goodput of \SI{73.9}{\%}.}
    \label{fig:dynamic:spmac}
  \end{subfigure}

  \caption{ Dynamic MAC aggregation can quickly adapt to varying channels and thus yield significant authenticated goodput improvements for short DTLS~1.3 transmission even under changing channel conditions.}
  \label{fig:dynamic}
\end{figure*}

\subsubsection{Goodput}

We show the achieved goodput in Figure~\ref{fig:static:goodput}.
These results validate the methodology from Section~\ref{sec:eval:estimation} and Figure~\ref{fig:goodput}.
Thus, MAC aggregation can indeed improve goodput significantly on realistic wireless channels.
For example, the authenticated goodput for 10-byte payloads achieved by traditional MAC nearly doubles from \SI{32.17}{\%} to \SI{61.34}{\%} with \code{Agg(16)}.
These results confirm the potential of MAC aggregation for concrete scenarios and confirm theoretical and simulative results from the past in a real deployment.

\subsubsection{Authentication Delays}

A valid concern about MAC aggregation schemes is authentication delay.
After a packet is received, the authenticity of the payload cannot always be immediately verified.
As discussed in Section~\ref{sec:design:interface}, our integration of MAC aggregation into DTLS~1.3 offers two ways to deal with this delay.
Either any not yet authenticated data can be buffered by the DTLS layer and forwarded to the higher communication layer once authenticity is guaranteed.
Alternatively, progressive authentication~\cite{2020_armknecht_promac} can be employed to optimistically push data to the higher layer and raise an alert in the unlikely case that a falsified MAC is detected retroactively.
For both approaches, it is important to understand \emph{how} MAC aggregation delays authenticity validation.

Therefore, we visualize this delay in Figure~\ref{fig:poc} for the 30-byte payload measurements for four MAC aggregation schemes, namely \code{Agg(8)}, \code{Agg(16)}, \code{R2D2(8,100)}, and \code{R2D2(8,50)}.
The y-axis shows the fraction of messages that have been authenticated to the given security level at a specific time.
The time is expressed on the x-axis in terms of sent follow-up messages after the initial message.
Thus, after five follow-up messages have been transmitted with \code{Agg(8)}, \SI{73.9}{\%} of messages have achieved a satisfactory security level of \SI{128}{bit} or higher (\tikz\draw[128bit,fill=128bit] (0,0) circle (.6ex);) and \SI{26.1}{\%} of messages completely lack any authenticity (\tikz\draw[0bit,fill=0bit] (0,0) circle (.6ex);), as can be seen in Figure~\ref{fig:poc:a}.
We see that \agg has lower authentication delays than \rd and that either a message is fully authenticated or not at all.

\rd behaves differently as data is progressively authenticated.
Thus, upon reception, all messages are already authenticated to a minimal level.
However, the same mechanisms that enable this progressiveness also lead to a longer time to reach full authenticity.
Thus, \rd is advantageous in scenarios where authentication delay is less important or if data can be processed with minimal security guarantees, while quick retrospective detection of manipulations is expected.

Overall, MAC aggregation leads to authentication delays that can be managed in different ways.
These delays differ vastly across aggregation schemes, so an educated choice is advised for a given deployment.

\subsection{Dynamic Parameter Selection}

In Section~\ref{sec:design:dynamic}, we presented a mechanism to dynamically adapt the parameters of the MAC aggregation scheme to current channel conditions.
Now, we investigate how conventional DTLS~1.3 connections compare to dynamic MAC aggregation with \agg and \rd.
To gather comparable results, we need a way to repeatedly alter the wireless channel conditions.

In our experiment, we want to change channel conditions based on movement as it occurs from mobile robots in industrial scenarios.
To mimic this behavior, we mount the sending Zolertial RE-Mote on a \emph{Dreame D9 Max} vacuum robot.
This robot is programmed to slowly move from a distance of \SI{1.9}{m} to a distance of \SI{13.2}{m} in a zigzag pattern over the course of \SI{15}{\minute}.
Afterwards, the robot drives straight back to the starting point in \SI{1}{\minute}.
For each MAC scheme, the robot drives this pattern twice with a \SI{5}{\minute}  pause in between, for a total evaluation duration of  \SI{36}{\minute}. 
We send 10 packets per second, each carrying one DTLS~1.3 frame with \SI{30}{\byte} of payload.

\begin{figure}[t]
  \includegraphics[width=\columnwidth]{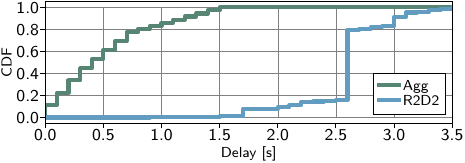}
  \caption{
    \agg leads to fast verification but a higher variance in delays than \rd.
  } 
  \label{fig:dynamic:delay}
\end{figure}

The results of our evaluation are shown in Figure~\ref{fig:dynamic}.
In Figure~\ref{fig:dynamic:vanilla}, we see the behavior of a conventional DTLS~1.3 connection.
The black line shows the running average of the authenticated goodput over the last 200 transmissions.
Likewise, the dark red line on the secondary y-axis shows the running average \ac{PER}.
Overall, the \ac{PER} remains low unless the robot is further away from the starting point between \SI{7}{\minute} and \SI{15}{\minute}, and between \SI{28}{\minute} and \SI{36}{\minute}.
However, over time, the achieved authenticated goodput remains relatively constant slightly below \SI{60}{\%}.
In total, the conventional DTLS~1.3 connection achieves a goodput of \SI{57.0}{\%}.

Figure~\ref{fig:dynamic:agg} now shows the performance of \agg with dynamically adjusted parameters.
Here, the bottom line does not only show the running average \ac{PER} on the second y-axis, but also the current MAC aggregation parameters through its color. 
We see that the parameters are regularly updated while the robot is in motion, selecting more conservative parameters as the channel worsens.
Thereby, the authenticated goodput surpasses \SI{80}{\%} on a good channel.
Meanwhile, the conventional DTLS~1.3 connection is also outperformed if the channel worsens, for an overall goodput of \SI{74.4}{\%}.

Finally, we show the results for \rd in Figure~\ref{fig:dynamic:spmac}.
In contrast to \agg, \rd can keep a high authenticated goodput of close to \SI{80}{\%} even when \acp{PER} reach \SI{2}{\%}.
Conversely, \rd is more severely impacted when \acp{PER} spike further.
During the spike after 10 minutes, we even see that channel switching to conventional DTLS~1.3 authentication for a short while. 
Despite these differences to \agg, \rd achieves a very similar overall goodput of \SI{73.9}{\%}.

Figure~\ref{fig:dynamic:delay} displays a CDF of the delays until data is fully authenticated for \agg and \rd in this scenario.
We again observe faster authentication for \agg, but more constant delays of mostly \SI{2.6}{s} for \rd.

Overall, MAC aggregation can significantly boost authenticated goodput, even in dynamic scenarios.
The raw performance of the different aggregation schemes does not vary significantly, such that secondary factors, like the shorter authentication delays of \agg or the support for longer payloads and resilience to selective jamming attacks by \rd, should determine the choice of the MAC aggregation scheme.

\section{Limitations and Future Research Potential}

Our evaluations show the potential of MAC aggregation in real-world deployments.
However, our evaluations also have limitations, and further optimizations are also possible.
For one, potential scenarios quickly increase in complexity with multiple devices, multihop communication, and variable-size traffic in a single DTLS session.
Our current evaluation setup does not cover these factors.
Thus, additional research is needed to understand the full potential of MAC aggregation, as all of these factors may influence performance and thus ultimately influence whether MAC aggregation adds value to a given scenario or not.

Additionally, MAC aggregation in DTLS~1.3 and other protocols may be further optimizable in practice.
The proposed integration process could be simplified or extended to offer more flexible configurations.
Additionally, the potential symbiosis of reliability-increasing strategies, \eg selective retransmissions or cooperative communication, and MAC aggregation is worth exploring, as MAC aggregation greatly benefits from low \acp{PER}.
Finally, the dynamic parameter selection procedure also offers potential to anticipate predictable changes to the channel quality (\eg because the distances between antennas change) or to reflect the dynamics of wireless channels better.

Exploring these future research directions thus potentially unlocks even better performance.
In that sense, the results presented in this paper can be understood as a lower bound of the potential of MAC aggregation.

\section{Conclusion}

MAC aggregation distributes the overhead of authentication tags over multiple messages to save valuable bandwidth, especially if transmitted payloads are short.
Recent results suggest that MAC aggregation can even improve authenticated goodput on wireless and, thus, lossy channels.
To the best of our knowledge, this paper presents the first real-world deployment of MAC aggregation for wireless communication.

Concretely, we show that MAC aggregation can be integrated into the DTLS~1.3 protocol while remaining standard-compliant and backward-compatible.
These efforts highlight the main challenges of deploying MAC aggregation in the real world, namely the definition of the interface to the upper communication layers and the process to (dynamically) negotiate MAC aggregation schemes and associated parameters.

Finally, we extensively evaluate MAC aggregation in DTLS~1.3 for an IEEE 802.15.4 wireless channel between two Zolertia RE-Mote boards running Contiki-NG.
We show that MAC aggregation indeed improves authenticated goodput significantly in many scenarios, more than doubling it for short payloads of only a few bytes.
Meanwhile, this process also leads to significant energy savings and can dynamically and efficiently adapt to changing channel conditions, even falling back to full-sized MACs if \acp{PER} increases excessively.
Overall, MAC aggregation can thus result in a much more efficient use of wireless communication channels.

\section*{Acknowledgments}

This work was partially funded by the Federal Ministry of Research, Technology and Space (BMFTR) in Germany under the grant number 16KIS2251. The responsibility for the content of this publication lies with the authors.

\bibliographystyle{IEEEtranS}
\bibliography{references}

\end{document}

%% file: include.tex
\usepackage{xcolor}
\usepackage[detect-weight=true,detect-family=true]{siunitx}
\usepackage{xspace}

\usepackage{tikz}
\usetikzlibrary{positioning}

\usepackage{multirow}

\usepackage{textgreek}

\definecolor{darkgrey}{RGB}{80,80,80}
\definecolor{lightgrey}{RGB}{170,170,170}

\definecolor{brown}{HTML}{a52a2a}
\definecolor{darkcyan}{HTML}{0a888a}

\definecolor{sender}{RGB}{132,200,234}
\definecolor{receiver}{RGB}{62,206,90}
\definecolor{authenticator}{RGB}{241,177,69}

\usepackage{adjustbox}
\usepackage{booktabs}
\usepackage{wasysym}
\usepackage{tabularx} 
\usepackage{multirow}
\usepackage{arydshln} 
\usepackage[detect-all]{siunitx}

\usepackage{enumitem}
\newlist{trianglelist}{itemize}{1}
\setlist[trianglelist,1]{nosep,label=$\blacktriangleright$,align=right,itemindent=\parindent+\widthof{~}+\widthof{$\blacktriangleright$},labelsep=\widthof{~},labelwidth=\widthof{$\blacktriangleright$},leftmargin=0pt}

\usepackage{colortbl}

\usepackage{flushend}

\usepackage{mathtools}

\newcommand{\ie}{\textit{i.e.},\xspace}
\newcommand{\eg}{\textit{e.g.},\xspace}
\newcommand{\cf}{\textit{cf.}\xspace}

\newcommand{\code}[1]{\texttt{#1}}

\usepackage{tikz}
\newcommand\copyrighttext{
\footnotesize
\textcopyright{} 2025 IEEE.
Personal use of this material is permitted.
Permission from IEEE must be obtained for all other uses, in any current or future media, including reprinting/republishing this material for advertising or promotional purposes, creating new collective works, for resale or redistribution to servers or lists, or reuse of any copyrighted component of this work in other works.
DOI: 10.1109/ICNP65844.2025.11192339
}
\newcommand\copyrightnotice{%
\begin{tikzpicture}[remember picture,overlay]
\node[anchor=south,yshift=25pt] at (current page.south) {\parbox{\dimexpr\textwidth-\fboxsep-\fboxrule\relax}{\copyrighttext}};
\end{tikzpicture}
}
